\documentclass[aps,pre,twocolumn,groupedaddress]{revtex4-1}

\usepackage[utf8]{inputenc}
\usepackage{amsmath}
\usepackage{amsfonts}
\usepackage{amssymb}
\usepackage{graphicx}
\usepackage{textcomp}
\usepackage{epstopdf}

\DeclareMathSizes{10}{10}{6}{5}

\begin{document}

\title{Behavior of Supercooled Aqueous Solutions Stemming from Hidden
Liquid-Liquid Transition in Water}
\author{John W. Biddle}
\author{Vincent Holten}
\author{Mikhail A. Anisimov}
\email{anisimov@umd.edu}
\affiliation{Institute for Physical Science and Technology, and Department of Chemical
and Biomolecular Engineering, University of Maryland, College Park, Maryland
20742, USA}

\begin{abstract}
A popular hypothesis that explains the anomalies of supercooled water is the existence of a metastable liquid-liquid transition hidden below the line of homogeneous nucleation. If this transition exists and if it is terminated by a critical point, the addition of a solute should generate a line of liquid-liquid critical points emanating from the critical point of pure metastable water. We have analyzed thermodynamic consequences of this scenario. In particular, we consider the behavior of two systems, H$_{2}$O-NaCl and H$_{2}$O-glycerol. We find the behavior of the heat capacity in supercooled aqueous solutions of NaCl, as reported by Archer and Carter, to be consistent with the presence of the metastable liquid-liquid transition. We elucidate the non-conserved nature of the order parameter (extent of ``reaction" between two alternative structures of water) and the consequences of its coupling with conserved properties (density and concentration). We also show how the shape of the critical line in a solution controls the difference in concentration of the coexisting liquid phases.
%\vspace{0.25cm}
%supercooled water | polyamorphism | aqueous solutions | critical phenomena | liquid-liquid transition
\end{abstract}

\maketitle

\section{Introduction}

There is a fascinating idea, known as \textquotedblleft water's
polyamorphism\textquotedblright , that hypothesizes the existence and possible phase separation of two
alternative structures of different densities in supercooled liquid water \cite{Poole_1992,Mishima_1998a,Debenedetti_2003a,Debenedetti_2003b}. This hypothesized liquid-liquid coexistence, terminated by a critical point, is not directly accessible to bulk-water experiments because it is presumably located a few degrees below the line of homogeneous nucleation of ice \cite{Debenedetti_2003b,Mishima_1998a,Meadley_2014}. Fresh approaches to resolving the question of the existence of water's polyamorphism are especially desirable in view of conflicting reports on simulations in water-like models \cite{Xu_2005,Xu_2006,Liu_2009,Sciortino_2011,Limmer_2011,Liu_2012,Kesselring_2012,English_2013,Kesselring_2013,Limmer_2013,Palmer_2013,Holten_2013a,Poole_2013,Yagasaki_2014}.

If the hidden liquid-liquid transition exists in metastable water, the addition of a solute will generate critical lines emanating from the pure-water critical point \cite{Chaterjee_2006,Anisimov_2012}. Thermodynamic analysis of these metastable critical phenomena would be conceptually similar to what is used \cite{Anisimov_1995a,Anisimov_1995b,Povodyrev_1997,Abdulkadirova_2002,Anisimov_2004} near the well understood vapor-liquid critical point of a solvent upon addition a solute. Moreover, in many aqueous solutions, as well as simulated models, the temperature of homogeneous nucleation is shifted to lower temperatures upon addition of a solute \cite{Kanno_1977,Miyata_2005,Kumar_2008}, which may provide a new way to access the vicinity of the hypothesized liquid-liquid transition.  Figure 1 shows suggested phase behavior of supercooled aqueous solutions of sodium chloride, in which the hypothetical liquid-liquid transitions between high-density liquid (HDL) and low-density liquid (LDL) are hidden by homogeneous nucleation.  Such behavior is also supported by experiments on the melting lines of metastable ice polymorphs in aqueous solutions of lithium chloride \cite{Mishima_2011} and by simulations of the TIP4P water model upon addition of sodium chloride \cite{Corradini_2010JCP,Corradini_2011JPCB}.
 
Archer and Carter \cite{Archer_2000} measured the heat capacity of pure water and aqueous NaCl solutions at ambient pressure and temperatures down to 236 K for pure water and down to 202 K in solutions.  They found a dramatic suppression of the heat-capacity anomaly upon addition of NaCl.  Archer and Carter have interpreted their results as evidence against the existence of the liquid-liquid transition in water.  On the contrary, we find the peculiar behavior of the heat capacity in metastable aqueous solutions of NaCl \cite{Archer_2000} to be in agreement with the hypothesis of a liquid-liquid transition and liquid-liquid critical point.  A suggested phase diagram for supercooled aqueous solutions of sodium chloride is shown in Fig. 1.

\begin{figure}
\includegraphics[width=0.49 \textwidth]{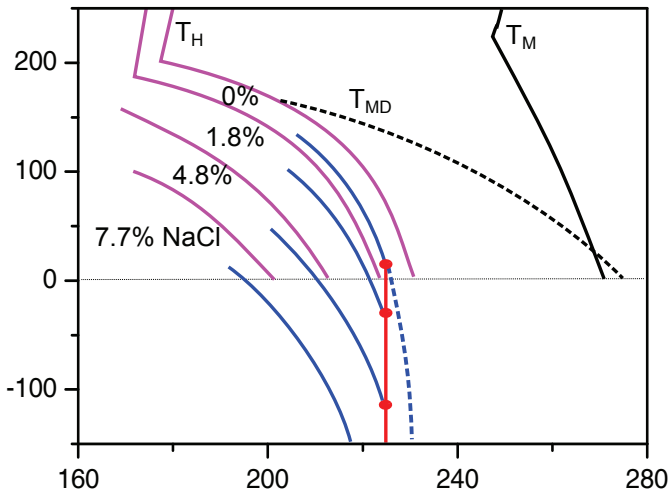}
\caption{Suggested phase diagram for supercooled aqueous solutions of sodium chloride exhibiting liquid-liquid transitions hidden by homogeneous ice formation.  Solid pink curves show homogeneous ice formation as obtained by Kanno and Angell \cite{Kanno_1977}.  $T_\textrm{M}$ labels the equilibrium melting temperatures of pure water.  The liquid-liquid transitions between HDL and LDL for various concentrations of NaCl are shown by blue curves.  The hypothesized critical line in the solution is shown by red, and the Widom line for pure water is shown by dashed blue.  The location of the critical point in pure water is shown as predicted by Holten and Anisimov \cite{Holten_2012b} ($T{_\textrm{c}}$=227.4~K, $P_{\textrm{c}}$=13.5~MPa). Upon cooling at constant composition, the phase transition line in the solution is split into two curves.  The blue curves correspond to the appearance of the first drop of LDL; the branches corresponding to the disappearance of the last drop of HDL are not shown.  $T_{\textrm{MD}}$ indicates the curve of maximum density in pure water.}
\end{figure}

Murata and Tanaka have reported direct visual observation of a liquid-liquid transition in supercooled aqueous solutions of glycerol \cite{Murata_2011}.
They have argued that the formation of a more stable liquid phase in this solution may occur by two alternative types of kinetics: nucleation and spinodal decomposition.  They have also claimed that the transition is mainly driven by the local structuring of water rather than of glycerol, suggesting a link to the hypothesized liquid-liquid transition in pure water.   However, they did not observe two-phase coexistence, leading them to claim that the transition is ``isocompositional" and the nucleation and spinodal decomposition occurs ``without macroscopic phase separation."

In this paper, we analyze the thermodynamic consequences of the existence of liquid-liquid transitions in supercooled aqueous solutions stemming from the liquid-liquid transition in pure water. Unlike liquid-liquid phase separation in binary solutions caused by non-ideality of mixing between two species \cite{Guggenheim_1949,Prigogine_1954}, the offspring of the liquid-liquid transition in pure water are begotten of the non-ideality of mixing between two alternative structures of water. We show that the behavior of these solutions is controlled by the shape of the critical line emanating from the critical point of pure water and by the thermodynamic path along which the transition is approached. We elucidate the nature of the scalar, non-conserved order parameter in supercooled water and aqueous solutions, its coupling with conserved properties such as density and concentration, and the character of nucleation and spinodal decomposition, which can occur with or without phase separation, depending on the thermodynamic path.

\section{Theory}
\subsection{Formulation of the Model: Two-Structures in Liquid Water}
Liquid--liquid phase separation in water can be elegantly explained if water is viewed as a mixture of two interconvertible structures, involving the same molecules, whose ratio is controlled by ``chemical-reaction" equilibrium \cite{Bertrand_2011}. The existence of two structures does not necessarily mean that phase separation will occur \cite{Holten_2013a,Tanaka_1999,Tanaka_2000JCP,Tanaka_2011FD}.  However, if the mixture of two structures is sufficiently non-ideal, a positive excess Gibbs energy of mixing could cause phase separation \cite{Holten_2012b,Holten_2014}. 

X-ray scattering \cite{Nilsson_2012} and spectroscopy experiments \cite{Taschin_2013} are consistent with the existence of a bimodal distribution of molecular configurations in water.  Furthermore, the existence of two different forms of liquid water is supported by the recent observation of two different glass transitions in water \cite{Amann-Winkel_2013}.

We assume that liquid water at low temperatures can be described as a mixture of a high-density structure A and a low-density structure B. Structure B is characterized by a hydrogen bond network similar to that in ice, with each molecule surrounded by four nearest neighbors.  In structure A, each molecule has up to two more nearest neighbors. It is important to note that both LDL and HDL are mixtures of the two structures A and B.  The fraction of molecules that form structure B in either liquid state is denoted by $%
\varphi $, and is controlled by the ``reaction" 
\begin{equation}
\text{A}\rightleftharpoons \text{B}.  \label{eq:reaction}
\end{equation}

The molar Gibbs energy $G$ is given by
\begin{equation}
G=(1-\varphi )\mu _{\mathrm{A}}+\varphi \mu _{\mathrm{B}}=\mu _{\mathrm{A}}+\varphi \mu _{\mathrm{BA}}
\end{equation}
where $\varphi $ is the mole fraction of structure B, $\mu _{\mathrm{A}}$ and $\mu_{\mathrm{B}}$ are the chemical potentials of A and B. The field variable conjugate to $\varphi$ is $\mu _{\mathrm{BA}}=\mu _{\mathrm{B}}-\mu_{\mathrm{A}}$. For the molar Gibbs energy we adopt an expression \cite{Holten_2012b} that accounts for the non-ideality of mixing in a simple symmetric form:
\begin{align}
&G=G_{\mathrm{A}}+\varphi G_{\mathrm{BA}}+ \nonumber \\ 
&RT\left[ \varphi \ln \varphi
+(1-\varphi )\ln (1-\varphi )+W\varphi (1-\varphi )\right] ,  \label{eq:G}
\end{align}
where $G_{\mathrm{A}}$ is the Gibbs energy of pure structure A, $G_{\mathrm{BA}%
}=G_{\mathrm{B}}-G_{\mathrm{A}}$ is the difference in Gibbs energies between the pure structures, $T$ is the temperature, and $W$, the measure of the nonideality of mixing, is generally a function of temperature and pressure.

The condition of chemical reaction equilibrium, 
\begin{equation}
\left( \frac{\partial G}{\partial {\varphi }}\right) _{T,P}=0,
\label{eq:dGdx}
\end{equation}%
defines the equilibrium fraction of $\varphi _{\mathrm{e}}$ of structure B, the extent or degree of reaction \cite{Guggenheim_1949,Prigogine_1954}.

\subsection{Nature of the Order Parameter and Classes of Universality}
The line of liquid-liquid transitions and the Widom line \cite{Bertrand_2011,Holten_2012b} (the smooth continuation of the transition line into the one-phase region, shown in Fig. 1) satisfy the condition 
\begin{equation}
\mathrm{ln}K = -\frac{G_{\textrm{BA}}}{RT} = 0, \label{eq:h10}
\end{equation}
where $K(T,P)$ is the equilibrium constant of the reaction. In the theory of phase transitions \cite{Fisher_1983}, the condition (5) corresponds to zero ordering field $h_1$ conjugate to the order parameter $\phi_1 = \varphi - \frac{1}{2}$ \cite{Holten_2012b}. In this theory, the two-phase region can be treated as the analogue of the spontaneously ordered state while other regions are analogues of states with non-zero ordering field.  Correspondingly, the order parameter is zero along the Widom line.  The order parameter spontaneously emerges in the two-phase region upon crossing the critical point and is also non-vanishing when it is induced by non-zero ordering field. We also note that HDL and LDL, like other fluids, possess continuous translational symmetry.  Thus the first-order transition between HDL and LDL is not accompanied by global symmetry-breaking.  Instead, this transition occurs upon a change in sign of the ordering field, $h_1 \propto \mathrm{ln}K$, across the liquid-liquid transition line.

The extent of reaction is a scalar, non-conserved physical property, meaning that its excess at a certain location is not necessarily compensated by a corresponding depletion elsewhere. Examples of non-conserved order parameters include magnetization in ferromagnets and degree of orientational order in liquid crystals. Thermodynamics of phase transitions with conserved and non-conserved order parameters can be quite similar. For example, all fluids near their critical points have a scalar, conserved order parameter (density and/or concentration).  Nevertheless, they belong to the same thermodynamic universality class as anisotropic (``Ising") ferromagnets near their Curie points, for which the order parameter is not conserved.  \cite{Fisher_1983}. However, fluids and Ising ferromagnets belong to fundamentally different universality classes in dynamics. When a system relaxes to equilibrium, a conserved order parameter obeys diffusion dynamics (its rate is space-dependent), while a non-conserved parameter equilibrates according to relaxation dynamics (the rate is space-independent) \cite{Hohenberg_1977}. This difference affects all dynamic phenomena, including spinodal decomposition and sound propagation.

However, there is an important feature of the extent of chemical reaction as the order parameter, which affects both dynamics and thermodynamics. This is a coupling of the non-conserved order parameter with conserved properties, such as density and energy.  This coupling is controlled by two coupling constants: $\lambda$, the heat of reaction (1) and the slope of the liquid-liquid transition line in the $(T,P)$ plane, $dP/dT = \Delta S / \Delta V$,  with $\Delta S \propto \lambda \Delta\varphi$  and $\Delta V$ are the changes in entropy and volume, respectively.  The dynamics of water and aqueous solutions near the liquid-liquid transition will be controlled by the competition between the rates of diffusion and relaxation.

In this work, we consider only the mean-field approximation of the two-structure model given by equation (\ref{eq:G}).  The effects of fluctuations have been addressed in Refs. \cite{Holten_2012b,Holten_2014}.  Fluctuations lead to non-analytic behavior of thermodynamic properties in the immediate vicinity of the critical point and cause a small shift in the critical parameters, however they do not qualitatively change the results presented here.

\subsection{Aqueous Solutions: Offspring of Water's Polyamorphism}
\subsubsection{Isomorphism}
There is a well-developed approach to treating the thermodynamics of mixtures near their critical points, known as ``isomorphism"  \cite{Wang_2008}.  Based on an examination of the stability criteria in fluids, it has been postulated that upon the addition of solute, the form of the equation of state remains unchanged under the condition of constant thermodynamic fields, including chemical potentials \cite{Griffiths_1970,Saam_1970,Anisimov_1971,Anisimov_1991,Anisimov_1995a,Anisimov_1995b,Wang_2008}.  

The molar Gibbs energy of a binary system is expressed through the
chemical potentials of the two components, solvent and solute, $\mu_{1}$
and $\mu_{2}$ as
\begin{equation}
G=(1-x)\mu _{1}+x\mu _{2}=\mu _{1}+x(\mu _{2}-\mu _{1}),
\end{equation}%
where $x$ is the mole fraction of solute and $\delta =$ $\left( \partial
G/\partial x\right) _{T,P\text{ }}=\mu _{2}-\mu _{1}$ is the thermodynamic field conjugate
to $x$, and
\begin{equation}
dG=VdP-SdT+\delta dx.
\end{equation}
In the theory of isomorphism, the chemical potential of the solvent in solution, $\mu_1 = G - x \delta$, which is the same in the binary-fluid coexisting phases, replaces the concentration-dependent Gibbs energy as the relevant thermodynamic potential such that
\begin{equation}
d\mu_1=VdP-SdT+xd\delta  \label{eq:chempotsolvent}
\end{equation}

There are two alternative cases of fluid-fluid separation in a binary solution. One is caused by non-ideality of mixing between the two species.  The other is  the offspring of a transition in the pure solvent.  The former case is typical for liquid-liquid separation in weakly compressible binary solutions, while the latter case is observed as fluid-fluid transitions stemming from the vapor-liquid transition in the pure solvent.  For the second case, the mixing of the two species in the solution does not need to be non-ideal, as the phase separation in the solution is a continuation of the phase-separation in the pure solvent \cite{Angell_1997}.  Kurita \emph{et al.} reported on a liquid-liquid phase transition in binary solutions of triphenyl phosphite with organic solutes such as diethyl ether or ethanol \cite{Kurita_2008}.  This transition stems from separation of pure triphenyl phosphite into liquid and amorphous states.  We model liquid-liquid transitions in supercooled aqueous solutions as instances of this latter case.

%It is also according to this latter case that we model the liquid-liquid transition supercooled aqueous solutions.

The stability criterion in fluid mixtures can be written in a form convenient for the latter case:
\begin{equation}
\left( \frac{\partial P}{\partial V}\right) _{T,\delta }=\left( \frac{\partial P}{\partial V}\right) _{T,x}+\left( \frac{\partial P}{\partial x}\right)_{T,V}^{2}\left( \frac{\partial x}{\partial \delta }\right) _{T,V} \leq 0.
\label{eq:stability2}
\end{equation}
%Along the two branches of the spinodal (the absolute stability limit of the corresponding liquid phase) in pure water, the isothermal compressibility and the isobaric heat capacity diverge. However, in a binary fluid both the compressibility and isobaric heat capacity at constant composition are generally finite.

We assume that the form of the isomorphic thermodynamic potential, which is the chemical potential of water in solution $\mu_1$, is the same as that of the Gibbs energy of pure solvent (water) given by Eq. (\eqref{eq:G}):

\begin{align}
&\mu_1=\mu_{\mathrm{A}} + \varphi\mu_{\textrm{BA}} = \mu _{1\mathrm{A}}+\varphi \mu _{1\mathrm{BA}}+ \notag \\ &RT\left[ \varphi \ln
\varphi +(1-\varphi )\ln (1-\varphi )+W\varphi (1-\varphi )\right]
\end{align}
where $\mu _{1\mathrm{A}}$ is the chemical potential of pure state A, and $\mu
_{1\mathrm{BA}}=\mu _{1\mathrm{B}}-\varphi \mu _{1\mathrm{A}}$, is the
difference in the chemical potentials between the pure states. The only, but
essential, difference from the pure-solvent thermodynamics is that the critical parameters, $T_{\mathrm{c}}$ and $P_{\mathrm{c}}$, and the nonideality parameter $W$ are now functions of the chemical potential difference $\delta = \mu_2 - \mu_1$. 

We must note that the chemical potentials of the two components, solvent and
solute, are not the chemical potentials $\mu _{\mathrm{A}}$ and $\mu _{\mathrm{B}}$ of two alternative structures in water; $\mu_1$ and $\mu_2$ are controlled by
the concentration of solution and interactions between the solvent and
solute molecules.

\subsubsection{Implications of Constant Composition}

The calculation of the properties at constant composition, the derivatives of the Gibbs energy, require performing a Legendre transformation $G=\mu _{1}+\mu x.$  In addition, we use an approximation, called the critical-line condition \cite{Anisimov_1995b}, that requires along the critical line in solution
\begin{equation}
x = x_{\mathrm{c}} = e^{\delta /RT}.  \label{eq:clc}
\end{equation}
When a critical line emanates from the critical point of the pure solvent, the principal thermodynamic property that controls the behavior of solutions at constant composition is the so-called Krichevskii parameter defined in the dilute-solution limit as \cite{Levelt-Sengers_1991}
\begin{equation}
\mathcal{K}=\lim_{x \to 0} \left( \frac{dP}{dx}\right) _{\mathrm{c,cxc}}=\frac{dT_{\mathrm{c}}}{dx}\left[ \frac{dP_{\mathrm{c}}}{dT_{\mathrm{c}}}-\left( \frac{dP}{dT}\right) _{\mathrm{c,cxc}}\right].  \label{eq:Krichevskii}
\end{equation}
%\begin{equation}
%\mathcal{K}=\lim_{x \to 0} \left( \frac{dP}{dx}\right) _{\mathrm{c,cxc}}=\frac{dP_{\mathrm{c}}}{dx}-\frac{dT_{\mathrm{c}}}{dx}\left( \frac{dP}{dT}\right) _{\mathrm{c,cxc}}.  \label{eq:Krichevskii}
%\end{equation}
where $\left( dP/dT\right)_{\mathrm{c,cxc
}}$ is the slope of the line of liquid-liquid coexistence at the critical point of pure solvent. The derivatives $dT_{\textrm{c}}/dx$ and $dP_{\textrm{c}}/dT_{\textrm{c}}$ determine the initial slopes of the critical line in the ($T,P,x$) space.  Thus, since in the limit of the solvent critical point $(dP/dx)_{\textrm{c,cxc}} = (\partial P/\partial x)_{T,V}$, the absolute stability limit (spinodal) can be formulated through the Krichevskii parameter as
\begin{equation}
\left( \frac{\partial P}{\partial V}\right)_{T,\delta }=\left( \frac{\partial P}{\partial V}\right) _{T,x}+\mathcal{K}^{2}\left( \frac{\partial x}{\partial \mu }\right) _{T,V}=0.  \label{stability3}
\end{equation}
Along the critical line the inverse compressibility at constant composition vanishes in the pure solvent limit as
\begin{equation}
\left( \frac{\partial P}{\partial V}\right) _{T,x}=-x\mathcal{K}^{2}\rightarrow 0.
\end{equation}
Note that if the fluid phase separation in solutions stems from the transition in pure solvent, the stability criterion of a pure fluid  smoothly transforms into the stability criterion of a solution.

If the solute dissolves more favorably in the higher-pressure phase, then le Chatelier’s principle indicates that the phase transition at a given temperature will move to lower pressures, and \textit{vice versa}.  The direction in which the phase transition pressure moves at constant temperature, positive or negative, indicates the sign of the Krichevskii parameter.  This does not necessarily mean that the sign of $dP_\textrm{c}/dx$ determines the sign of the Krichevskii parameter, as this derivative is influenced both by the movement of the transition line in the $(T,P)$ plane and the movement of the critical point along that line.

In binary fluids, the critical point becomes a critical line and the phase-transition line becomes a surface of two-phase coexistence in the ``theoretical" $(T,P,\delta )$ space.  However, in the ``experimental" $(T,P,x)$ space the behavior of thermodynamic properties evaluated at constant composition will, in general, be different from that of the corresponding properties in the pure solvent and from that of the isomorphic properties in solutions (evaluated at constant chemical-potential difference $\delta$). Remarkably, the nature and magnitude of this
difference depend primarily on the value of the Krichevskii parameter \cite{Anisimov_1995a,Anisimov_1995b}.

In particular, the concentration gap in the $\left( T,x\right) $ plane at constant temperature can be found from Eq. (\ref{eq:chempotsolvent}) as
\begin{equation} \label{eq:C-C}
\left( \frac{dP}{d\delta}\right) _{T,\mathrm{cxc}}=\frac{\Delta x}{\Delta V},
\end{equation}
where $\Delta V$ is the difference in volume of the coexisting phases.  In the dilute-solution approximation, $dx/d\delta = x/RT$, so the concentration gap at the first-order transition and constant pressure can be evaluated through the Krichevskii parameter, $\Delta V$, and the slope of the transition line as
\begin{equation}\label{eq:xgap}
\Delta x \simeq -x\mathcal{K}\frac{\Delta V}{RT},
\end{equation}
where $x=x_{\textrm{c}}$ in accordance with the critical-line condition (\ref{eq:clc}).  Correspondingly, the temperature gap at constant composition can be evaluated as (see Appendix)
\begin{equation}
\Delta T \simeq x\mathcal{K}^{2}\frac{\Delta V}{RT}\left( \frac{dT}{dP}\right)_{\mathrm{c,cxc}}.  \label{eq:temperaturegap}
\end{equation}
In the solvent-critical-point limit, $\Delta V \propto x$ and the phase diagram develops a so-called ``bird's beak" where the concentration gap vanishes to first order in $x$ and the two branches of the transition merge with the same tangent \cite{Povodyrev_1997,Levelt-Sengers_1991}.

The above-described thermodynamics explains possible phase behavior of a supercooled aqueous solution with a critical line emanating from the pure solvent (water) critical point, as shown in Figs. 2 and 3. Only in a special case, when the the critical line and the liquid-liquid transition line merge with the same slope in the ($P,T$) plane, the Krichevskii parameter is zero, and the liquid-liquid transition in solution will be isocompositional. That case corresponds to the so-called critical azeotrope \cite{Anisimov_1995a,Anisimov_1995b}.  The case demonstrated in Figures 2 and 3 corresponds to a negative value of the Krichevskii parameter.  The sign of the Krichevskii parameter determines the partition of the solute between the coexisting phases.  The negative sign of the Krichevksii parameter means that HDL has a higher concentration of the solute.

\begin{figure}
\includegraphics[width=0.49\textwidth, trim = 0cm 12cm 0cm 0cm]{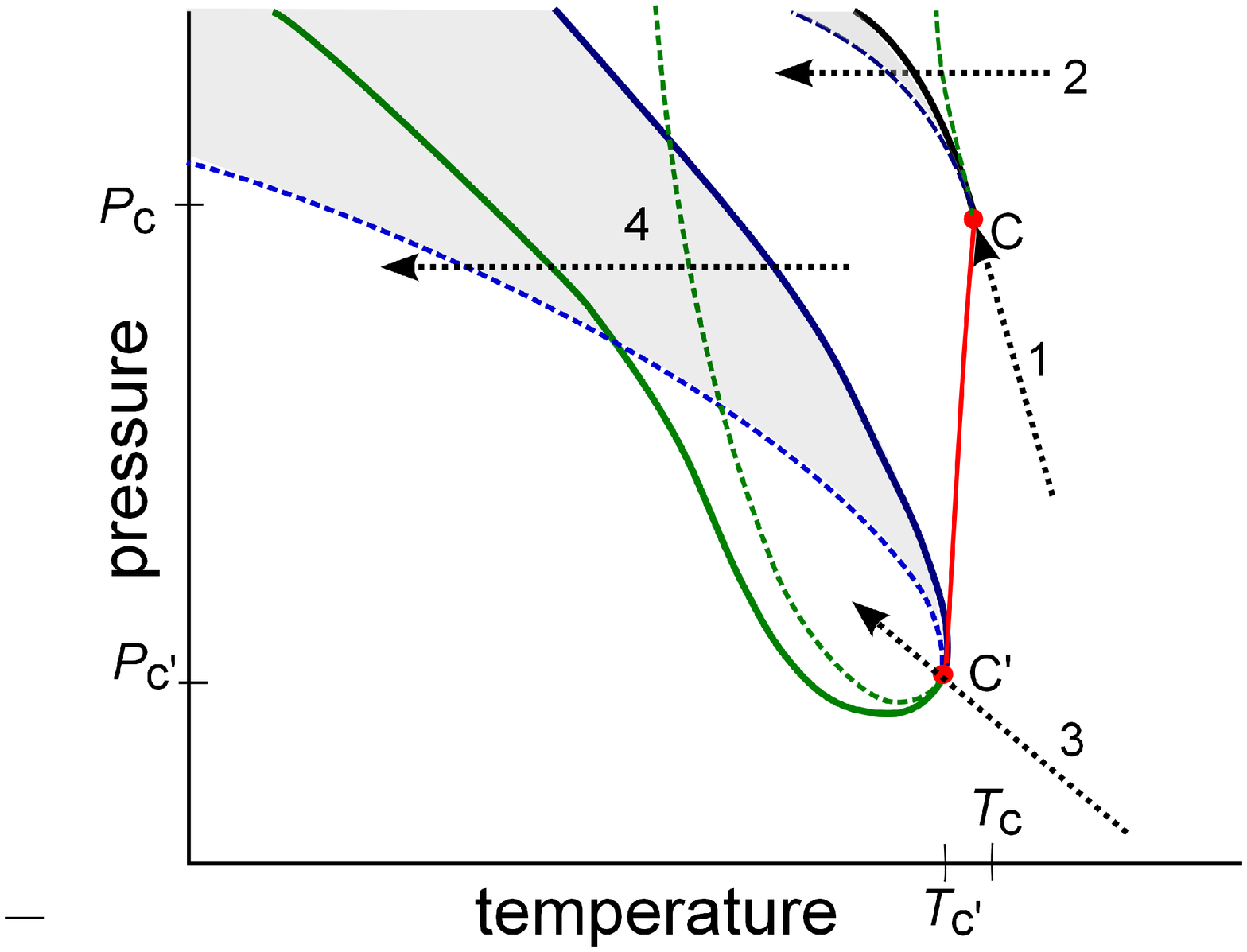}
\caption{ An example of phase boundaries at constant composition in a supercooled aqueous solution exhibiting a liquid-liquid transition between HDL and LDL.  The black curve is the liquid-liquid transition in pure water, terminated at the critical point C.  The critical line is shown by solid red with the critical point of the solution labeled C$'$. The blue curve shows the appearance of the first droplet of of LDL.  The green curve shows the disappearance of the last droplet of HDL.  The blue and green dashed curves are the thermodynamic stability limits of HDL and LDL, respectively. The shaded region shows where HDL forms by nucleation.  The dotted lines labeled 1,2,3, and 4 show different thermodynamic paths as explained in the text.}
\end{figure}

\begin{figure}
\includegraphics[trim = 0cm 0cm 0cm 0cm, width=0.49\textwidth]{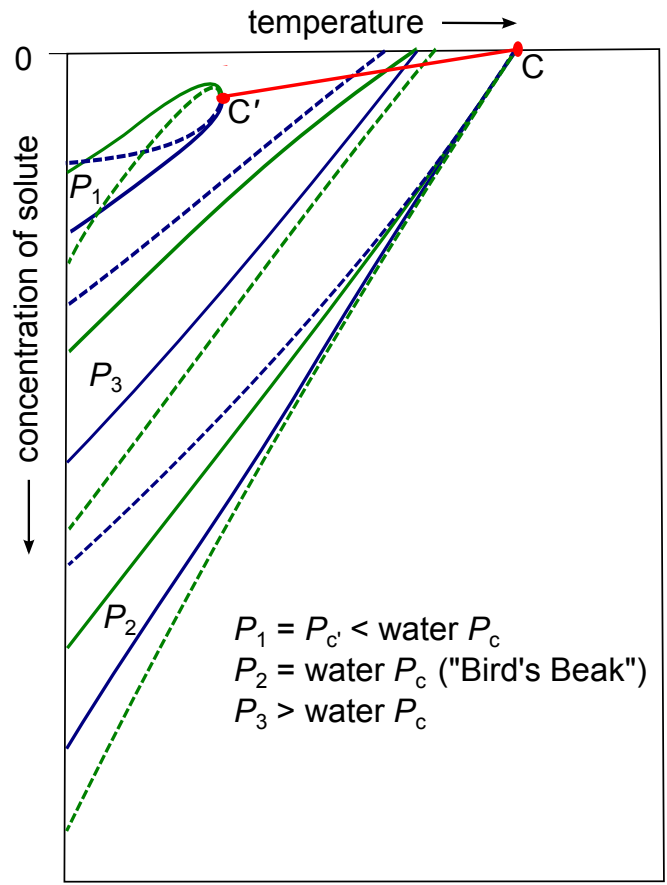}
\caption{Schematic $T$-$x$ diagram of a supercooled aqueous solution exhibiting a liquid-liquid transition between HDL and LDL.  The red line is the critical line with the critical point of pure water labeled C and the critical point of the solution at a certain concentration C'.  Solid blue and green lines show the coexistence between two phases, HDL and LDL, respectively.  Blue and green dashed lines show the thermodynamic stability limits of HDL and LDL, respectively.}
\end{figure}

The existence of phase separation in two-structure thermodynamics, caused by coupling of the order parameter with density and entropy, raises an interesting question on the path dependence of the character of spinodal decomposition in such systems. Conventionally, spinodal decomposition in fluids is observed along the critical isochore which, asymptotically close to the critical point, merges with the Widom line.  For this path, the final equilibrium state will be the two-phase coexistence between liquid and vapor. However, if a fluid, initially (for example) in the gaseous state, is quenched at constant pressure to the liquid state, the formation of the new equilibrium state may occur by two alternative mechanisms, either nucleation or spinodal decomposition, both without macroscopic phase separation. The same will be true for the liquid-liquid transition in water. This is illustrated in Fig. 2.  The conventional spinodal decomposition toward macroscopic phase separation will be observed upon quenching along paths 1 (pure water) and 3 (solution).  However, if the final equilibrium state is located in the shaded region between the spinodal (the absolute stability limit of the high-temperature liquid) and the phase transition line, the  new state will be formed by nucleation without macroscopic phase separation. If the final state is reached beyond the spinodal, the process will be similar to spinodal decomposition, but without macroscopic phase separation.  These events are illustrated in Fig. 2 by thermodynamic paths 2 and 4.

\section{Results}

\subsection{Suppression of Heat Capacity Anomaly in Aqueous Solutions of Sodium Chloride}

The experimental information on the thermodynamic properties of supercooled aqueous solutions of salts, in particular of NaCl, is very limited.  The available data are the isobaric heat capacity measurements of Archer and Carter \cite{Archer_2000}, and the density measurements of Mironenko \textit{et al.} \cite{Mironenko_2001}, both at atmospheric pressure.  Archer and Carter observed that for small NaCl concentrations, upon lowering the temperature, the heat capacity increases in the supercooled region. As the salt concentration is increased, this anomalous rise in heat capacity moves to lower temperatures and decreases in magnitude, virtually disappearing for salt concentrations greater than 2 mol/kg.  Mironenko \textit{et al.} found that as the concentration of NaCl was increased, the density of the solution increased while the density maximum moved to lower temperatures \cite{Mironenko_2001}.  About forty years ago, Angell observed the suppression of the heat capacity anomaly in supercooled water upon addition of lithium chloride \cite{Angell_40yrsago}, qualitatively similar to the effects reported by Archer and Carter for sodium chloride \cite{Archer_2000}.  In light of what can be inferred about the movement of the locus of liquid-liquid phase transitions upon addition of NaCl, this behavior of the heat capacity is precisely what thermodynamics predicts if the anomaly in pure supercooled water is indeed associated with a liquid-liquid critical point.

Homogeneous ice nucleation in solutions of NaCl is shifted to lower temperatures as the concentration of salt increases \cite{Kumar_2008,Kanno_1977}, with the lines of homogeneous nucleation keeping nearly the same shape in the $(T,P)$ plane as in pure water \cite{Kanno_1977}. The kinks in the melting lines of metastable phases of ice in aqueous solutions of LiCl, observed by Mishima, suggest that the liquid-liquid transition also moves to lower temperatures and pressures as the salt is added, remaining just below the temperature of homogeneous nucleation for any given concentration of solute \cite{Mishima_2011}.  Mishima has also observed that the transition in amorphous water between the HDA and LDA phases moves to lower pressures upon addition of LiCl \cite{Mishima_2007}.

Hypothesized phase behavior of supercooled aqueous solutions of sodium chloride showing the liquid-liquid transitions between HDL and LDL is presented in Fig. 1. The location of the critical point in pure water along the liquid-liquid transition is uncertain; in Fig. 1 it is shown according to the recent estimate, about 13 MPa, of Ref. \cite{Holten_2012b}.  However, according to the analysis of Ref. \cite{Holten_2012b}, one can currently only say that the critical pressure is smaller than 30~MPa, and could even be negative.   Above the lines of homogeneous ice formation, negative pressures are experimentally accessible and correspond to doubly metastable liquid water, with respect to both the solid and vapor states \cite{Pallares_2014}.   Liquid-liquid transitions at these pressures are an intriguing possibility \cite{Tanaka_1996,Brovchenko_2005,Stokely_2010,Meadley_2014}.

Simulations on the TIP4P\cite{Corradini_2011JPCB} and mW \cite{Le_2011} models of water suggest that hydrophilic solutes dissolve more easily in HDL than in LDL, the tetrahedral structure of which they tend to disrupt.  This further corroborates the hypothesis that the liquid-liquid transition and Widom line will move to lower pressures (at constant temperature) and to lower temperatures (at constant pressure) as the concentration of salt increases. Corradini and Gallo examine the slope of the liquid-liquid phase transition line and the position of the liquid-liquid critical point in TIP4P water at several concentrations of NaCl \cite{Corradini_2011JPCB}.  From these results we can estimate the derivatives in Eq. \eqref{eq:Krichevskii}
as follows: $dT_{\textrm{c}}/dx =$ 770 K, $dP_{\textrm{c}}/dT_{\textrm{c}}=-13.1$~MPa/K, and $(dP/dT)_{cxc} = -3.1 \hspace{1mm} \textrm{MPa/K}$, yielding a value of -7700~MPa for the Krichevskii parameter in this water model. 
Such a large magnitude of the Krichevskii parameter indicates that the critical anomalies will be greatly suppressed even for small concentrations of NaCl, and its sign indicates that NaCl dissolves better in HDL than in LDL.

As can be seen in Fig. 4, our equation of state is in qualitative agreement with simulation studies on the TIP4P model of water.  Both our equation of state and the simulations of Corradini and Gallo \cite{Corradini_2011JPCB} display a large, negative value of the Krischevksii parameter, driven primarily by the movement of the critical point to lower pressures.  Corradini and Gallo also find a slight increase in the critical temperature as NaCl is added, and they find a smaller slope for the LLT.  A re-scaling of the transition line obtained for the TIP4P model to match the slope of the transition line in our equation of state suggests an almost vertical critical line in real NaCl solutions (Fig. 4). A vertical critical line is adopted in our equation of state and, as shown below, is also supported by further analysis of the heat capacity data.

\begin{figure}[h!]
\includegraphics[trim = 0cm 0cm 0cm 0cm, width = 0.49\textwidth]{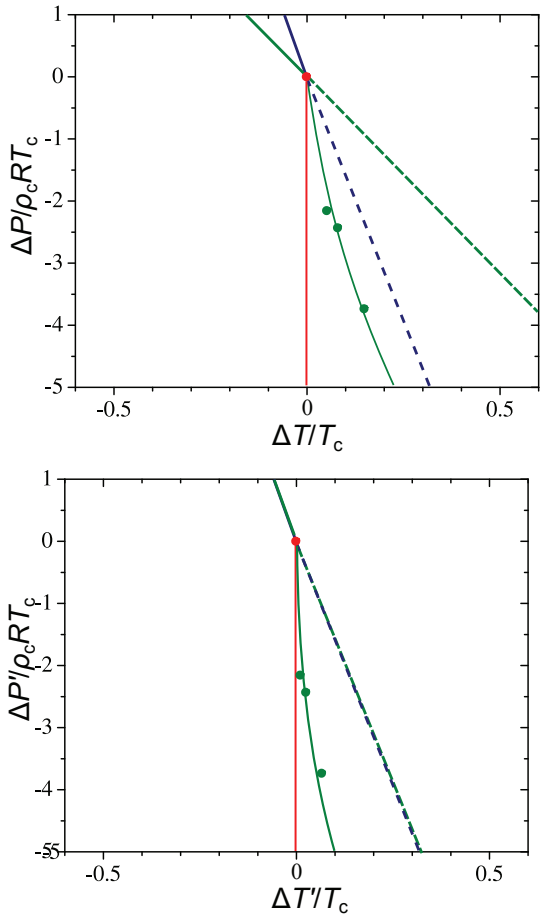}
\caption{ A comparison of our equation of state with the simulation results of Corradini and Gallo for the TIP4P model \cite{Corradini_2011JPCB}.  Because the systems have very different critical pressures, the features are presented in terms of difference from the critical point in variables reduced by the critical parameters of the system, as indicated.  The blue solid line and blue dashed line show linear approximations of the LLT and Widom line, respectively, for our equation of state.  The green solid line and green dashed line show a linear approximation of the LLT and Widom line, respectively as reported in Ref. \cite{Corradini_2011JPCB} for the TIP4P model.  Green circles show the critical points calculated at different mole fractions of NaCl in simulation with the thin green line as a guide to the eye, whle the red line shows the critical line for our equation of state.  The critical point of H2O is shown as a red circle. Top: our data and that of Corradini et al.  Bottom: data of Corradini et al. re-scaled so that the the LLT has the same slope as in our equation of state.}
\end{figure}

Simulations, experiments on the metastable ices in aqueous LiCl, and experiments on the homogeneous nucleation in NaCl are thus in agreement that the locus of liquid-liquid transitions moves rapidly to lower temperatures and pressures upon addition of NaCl, yielding a negative Krichevskii parameter on the order of $10^3$~MPa.  In order to form a more precise estimate for our model, we take note of Mishima's evidence that in solutions of LiCl, the liquid-liquid transition remains just below the line of homogeneous ice nucleation as both move to lower temperatures and pressures \cite{Mishima_2011}. With a linear approximation for the curve comprising the locus of liquid-liquid phase transitions and the Widom line, the Krichevskii parameter can be calculated based on the movement of this curve, regardless how the critical point might move along it.  Thus, taking the behavior of the line of homogeneous nucleation as a proxy for that of the line of liquid-liquid phase transitions, Ref. \cite{Kanno_1977} gives a Krichevskii parameter of $\mathcal{K} =-2230~\textrm{MPa}$ and Ref. \cite{Kumar_2008} gives $\mathcal{K} = -2860~\textrm{MPa}$.  Within that range, the value that we adopt for the Krichevskii parameter makes only a small difference in the fit of the model to the data, and the slope of the critical line makes little difference provided that the dominant contribution to the Krichevskii parameter comes from the movement of critical point to lower pressures, as suggested by \cite{Corradini_2011JPCB}.  We find that a vertical critical line and a Krichevskii parameter of $\mathcal{K} = -2860~\textrm{MPa}$ provides the most accurate calculations of the heat capacity, and accordingly adopt these parameters.

Salts and sugars depress the temperature of maximum density in water \cite{Cawley_2006}. For dilute solutions of simple electrolytes such as NaCl, there is a linear relationship between the concentration of the solute and the and depression of the temperature of maximum density, a relationship known as Despretz's law \cite{Wright_1919,Despretz_1839,Despretz_1840}.  For those salinities at which data exist, our equation of state reproduces this phenomenon adequately and matches the experimental data, as shown in Fig. 5.

\begin{figure}[h!]
\includegraphics[trim = 0cm 0cm 0cm 0cm, width = 0.49\textwidth]{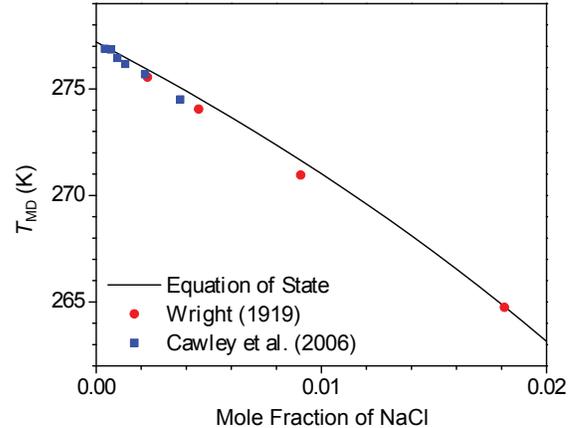}
\caption{Temperature of maximum density in aqueous solutions of supercooled water.  The black line shows the equation of state used in this work, while the red squares and blue circles show the measurements of Refs. \cite{Wright_1919} and \cite{Cawley_2006} respectively.}
\end{figure}

To calculate the isobaric heat capacity at constant
composition we use the thermodynamic relation between this experimentally available property and the ``theoretical" (isomorphic) heat capacity $C_{P,\delta }=T\left( \partial S/\partial T\right) _{P,\delta}:$ 
\begin{equation}
C_{P,x}=C_{P,\delta }-T\frac{\left( \partial x/\partial T\right) _{P,\delta }^{2}}{\left( \partial x/\partial \delta \right) _{P,T}},  \label{eq:legendre1}
\end{equation}
yielding
\begin{equation} \label{eq:cp}
\frac{C_{P,x}}{R}=\frac{T}{R}\left( \frac{\partial S}{\partial T}\right)_{P,x} = \hat{a}^{2}\frac{\chi_{1}}{1 + x(\phi_1\hat{\mathcal{L}}+\hat{\mathcal{K}})^2\chi_1}+B,
\end{equation}
where the background heat capacity $B$ is approximated as a polynomial function of $T$ and $x$, and $\hat{a} = (\rho_c R)^{-1} (dP/dT)_{\textrm{c,cxc}}$, $\hat{\mathcal{K}} = \mathcal{K}/\rho_c R T_{\mathrm{c}}$, $\chi_1$ is a strongly divergent susceptibility, and $\hat{\mathcal{L}} = \left(dP_{\textrm{c}}/dx\right)/\rho_{\mathrm{c}} R T_{\mathrm{c}}$ (see Appendix).

Equation (\ref{eq:cp}) describes the crossover of the heat capacity between two limits.  In the limit $x\to~0$ one recovers the expression for the heat capacity of pure water, diverging at the critical point as 
\begin{equation}
\frac{C_p}{R} = \hat{a}^2\chi_1 .
\end{equation}
As the solution critical point is approached, $\chi_1 \to \infty$, $\phi \to 0$, and the heat capacity approaches a finite value, growing with decreasing concentration:
\begin{equation}
\frac{C_P}{R} \to \frac{\hat{a}^2}{x\hat{\mathcal{K}}^2} + B.
\end{equation}

The large negative value of the Krichevskii parameter for this system, $\hat{\mathcal{K}}\simeq -30$ is mainly responsible for the significant suppression of the heat capacity anomaly even in dilute solutions of NaCl. The results of fitting Eq. (\ref{eq:cp}) to the experimental data of Archer and Carter are shown in Fig. 6.  To describe the heat capacity of NaCl solutions, we have used the mean-field version of the equation of state developed by Holten and Anisimov \cite{Holten_2012b} with the extrapolated mean-field value of the critical pressure in pure water, practically equal to atmospheric pressure.  The agreement between the theory and experiment is remarkable.

\begin{figure}[h!]
\includegraphics[width=0.49\textwidth]{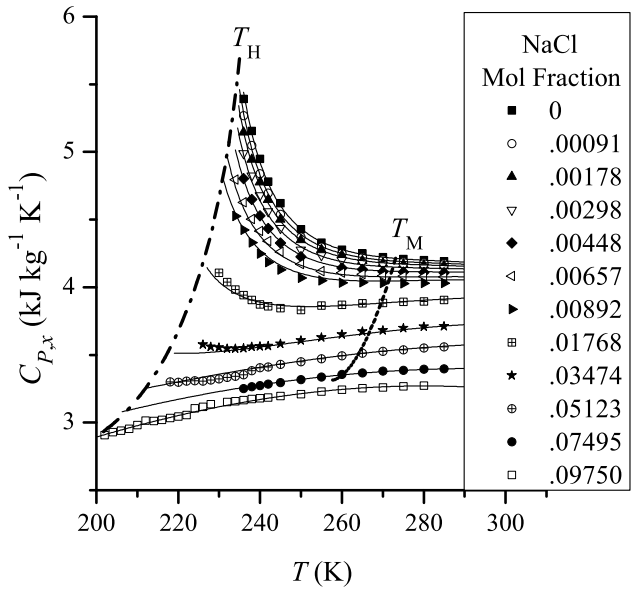}
\caption{Suppression of the anomaly of the heat capacity in aqueous solutions of sodium chloride.  Symbols: experimental data of Archer and Carter \cite{Archer_2000}.  Solid curves: predictions based on two-state thermodynamics.  Dashed curve shows the positions of the melting temperatures as given by the IAPWS formulation for saltwater \cite{IAPWS_2008}.  Dashed-dotted curve shows the temperatures of homogeneous ice formation \cite{Kanno_1977}.}
\end{figure}

\subsection{Liquid-Liquid Transition in Glycerol-Water}

Addition of glycerol lowers the temperature of homogeneous nucleation. Glycerol stabilizes the liquid state, because hydrogen bonding between water and glycerol increases the nucleation barrier for ice formation \cite{Murata_2011}. At mole fractions of glycerol $x \geq 0.135$, Murata and Tanaka have reported on phase transitions between two liquid states in the solution. They observed two alternative types of kinetics in the formation of the low-temperature liquid state: nucleation and spinodal decomposition. They have also found that the transition is mainly driven by the local structuring of water rather than of glycerol, suggesting a link to the hypothesized  transition between LDL and HDL in pure water. However, Murata and Tanaka have also claimed that the transition between two liquids in supercooled water-glycerol solutions is ``isocompositional," \textit{i. e.}, at the transition point, LDL and HDL have the same concentration of glycerol.  They also argue that the transition occurs without macroscopic phase separation.  Furthermore, they relate these putative features of the phase transition to the non-conserved nature of the order parameter.

We suggest an alternative interpretation of the experiments of Murata and Tanaka. As we have shown above, the HDL-LDL transition in aqueous solutions stemming from the transition in pure water cannot be isocompositional, except for the case of a special behavior of the critical line, yielding the Krichevskii parameter to be zero.  Moreover, it cannot take place without macroscopic phase separation if there exists a coupling between the order parameter and density and concentration. 

As suggested by Murata and Tanaka, In HDL glycerol molecules destabilize hydrogen bonding as pressure does in pure water, whereas in LDL cooperative inter-water hydrogen bonding and the resulting enhancement of tetrahedral order promote clustering of glycerol molecules. This suggests that the critical line emanating from the critical point of pure water continues down to negative pressures, while the critical temperature decreases.  Therefore, the difference in interaction of glycerol molecules with the two alternative liquid structures practically rules out the possibility that the critical point moves tangent to the phase transition line $(dP/dT)_{\textrm{c,cxc}}$.  Thus the coexisting phases will not have the same composition and the Krichevskii parameter will not be zero. Adopting the extrapolation of Murata and Tanaka for atmospheric pressure, the critical point of the solution will be found at $x \simeq 0.05$ and $T \simeq 225$~K, and locating the critical point of pure water at 13~MPa and 227~K as suggested in Ref. \cite{Holten_2012b}, we obtain from equation (\ref{eq:Krichevskii}) the Krichevskii parameter to be $\mathcal{K} \simeq -600~\mathrm{MPa}$.  The temperature gap for the transition at constant concentration, \textit{e. g.} $x = 0.165$ and atmospheric pressure can be evaluated from equation (\ref{eq:temperaturegap}).  The difference in the molar volumes $\Delta V/V_{\textrm{c}}$ of the coexisting phases can be estimated as about $0.05$ based on the distance between the transition at atmospheric pressure and the critical point at the same concentration of glycerol.  Then we find $\Delta T \simeq 5~\mathrm{K}$.

In light of this, it is unsurprising that Murata and Tanaka observed the formation of LDL alternatively by spinodal decomposition and by nucleation without observing macroscopic phase separation at $x = 0.165$.  As illustrated in Fig. 2, the transition should occur through spinodal decomposition if it takes place below the absolute stability limit of the HDL phase, and by nucleation and  growth if it takes place between the point where the last drop of HDL vanishes in the meta-stable state and the absolute stability limit.  The slow kinetics in supercooled water-glycerol and the narrow width $\Delta T$ of the two-phase region make this scenario worthy of consideration.  However, available experimental data of the phase behavior of supercooled glycerol aqueous solutions are still inconclusive. Other interpretations of the results reported by Murata and Tanaka \cite{Murata_2011}, in particular regarding the role of partial crystallization, might be considered. Further experimental studies of this system are highly desirable.

%\begin{figure}[h!]
%\includegraphics[trim = 0cm 0cm 0cm 0cm, width = 0.49\textwidth]{Glycerol7}
%\caption{An interpretation of phase behavior in supercooled aqueous solutions of glycerol based on observations of Murata and Tanaka.  The bottom plane of the graph is the cross-section at atmospheric pressure.   The surface of homogeneous ice formation is shown by pink.  C is the estimated location of the liquid-liquid critical point in pure water.  C$^{\prime}$ is the predicted location of the critical point in solution at atmospheric pressure. The light blue surface shows the appearance of the first droplet of LDL as the solution is cooled.  Solid blue curves are examples of isobars or isopleths along this surface.  The dashed blue curves are stability limits for HDL.  Solid green curves show where the last droplet of HDL disappears.  Between the green and blue lines, LDL and HDL coexist with different concentrations of glycerol.  The black circles mark observations of formation of LDL by spinodal decomposition.  The square shows an observation of formation of LDL by nucleation and growth at $x = 0.165$.}
%\end{figure}

\section{Conclusion}

Peculiar behavior of supercooled aqueous solutions may be an indication of metastable water's polyamorphism. Analysis of the scenario in which liquid-liquid transitions in binary solutions are offspring of the hypothesized transition between HDL and LDL in the solvent (pure water) show that 
thermodynamics imposes certain restrictions on the behavior of such binary solutions. In particular, the transition is generally accompanied by macroscopic phase separation due to the coupling between a non-conserved order parameter characterizing the difference in the structures of HDL and LDL and conserved properties, such as density and concentration. The width of the macroscopic phase separation and the change in the thermodynamic anomalies  is mainly controlled by the Krichevskii parameter,  a combination of the direction of the critical line emanating
from the pure-water critical point and the slope of the liquid-liquid transition in the ($P,T$) space. The critical anomalies shown by the response functions in the pure fluid will be suppressed when measured at constant composition in the solution.  The fact that the crossover behavior of the heat capacity in metastable aqueous solutions of NaCl is well described by this thermodynamics supports the idea of water's polyamorphism.

Unlike the well understood liquid-liquid phase separation in binary solutions caused by sufficient non-ideality of mixing between two species, the transitions springing from of the liquid-liquid transition in pure water are not driven by non-ideality of mixing between the solute and the solvent.  For example, nearly ideal mixtures of metastable H$_2$O and D$_2$O could manifest the critical line connecting the liquid-liquid critical points of these two species.  In this particular case, the only reason for liquid-liquid transition is sufficient non-ideality of mixing between two alternative structures in each species.

In solutions, upon a quench at constant pressure and constant overall composition, the formation of the new
equilibrium state may occur by two alternative mechanisms,
nucleation or spinodal decomposition, each either with or without macroscopic phase separation. This will depend on the path which is used to approach the equilibrium state and on the nature of the state. If the temperature gap of the transition is narrow and if the final equilibrium state is macroscopically homogeneous, both nucleation and spinodal decomposition will occur without macroscopic phase separation.

\section{Acknowledgments}

Acknowledgment is made to the donors of the American Chemical Society Petroleum Research Fund for support of this research (Grant No. 52666-ND6). Research of V.H. was partially supported by the National Science Foundation (Grant No. CHE-1012052). M.A.A highly appreciates fruitful discussions with C. A. Angell, I. Abdulagatov, S. Buldyrev, P. Debenedetti, P. Gallo, O. Mishima, V. Molinero, J. V. Sengers, H. E. Stanley, and H. Tanaka.

\section{Appendix: Scaling Fields and the Krichevskii Parameter}
\subsection{Heat Capacity at Constant Composition}
In the theory of critical phenomena, the thermodynamic potential can be separated into a regular background part and a critical part.  The critical part of the potential is associated with the dependent scaling field $h_3$, which can be expressed in terms of two independent scaling fields: the ordering field $h_1$ and the second, ``thermal" field $h_2$.  
\begin{equation}
h_3 = \phi_1 dh_1 + \phi_2 dh_2,
\end{equation}
where $\phi_1$ is the order parameter and $\phi_2$ is the second (weakly fluctuating) scaling density.

When the molar Gibbs energy $G(T,P)$ is used as the thermodynamic potential, the independent scaling fields can be expressed in linear approximation as combinations of the temperature $T$ and pressure $P$, expressed as \cite{Holten_2012b,Fuentevilla_2006,Bertrand_2011,Holten_2012a}
\begin{eqnarray}
h_1 = a_1\Delta P + a_2\Delta T, \\
h_2 = b_1\Delta T + b_2\Delta P.
\end{eqnarray}
For pure water, we take
\begin{align}
a_1 &= \frac{1}{\rho_{\textrm{c}}RT_{\textrm{c}}} \hspace{5mm} a_2 = -\frac{1}{\rho_{\textrm{c}}RT_{\textrm{c}}}\left(\frac{dP}{dT}\right)_{\textrm{c,cxc}}, \\
b_1 &= 0 \hspace{5mm} b_2 = \frac{1}{\rho_{\textrm{c}}RT_{\textrm{c}}},
\end{align}
where $\left(dP/dT\right)_{\textrm{c,cxc}}$ is the slope of the phase transition line at the critical point. The condition $b_1=0$ corresponds to an entropy-driven phase separation \cite{Holten_2012b}.

In a two-component mixture there is an additional thermodynamic degree of freedom to consider, and the scaling fields should be generalized to \cite{Anisimov_1995a,Anisimov_1995b}
\begin{align}
h_1 = a_1\Delta P + a_2\Delta T + a_3\Delta \delta, \\
h_2 = b_1\Delta T + b_2\Delta P + b_3\Delta \delta.
\end{align}

According to the principle of critical-point universality, the dependent scaling field $h_3$ must depend on the independent scaling fields $h_1$ and $h_2$ in the same way for a mixture as for a pure fluid. Our approximation that the isomorphic Gibbs energy $\mu_1 = G-x\delta$ retains the same form in mixtures as in pure water entails that the coefficients in the scaling fields, $a_1$, $a_2$, $b_1$, and $b_2$ remain unchanged.

With respect to an arbitrary point on the critical line, the scaling fields can be expressed to linear order as 
\begin{align}
h_1 = a_1\Delta P + a_2\Delta T - \left(a_1\frac{dP_{\textrm{c}}}{d\delta}\Delta\delta + a_2\frac{dT_{\textrm{c}}}{d\delta}\Delta \delta\right), \\
h_2 = b_2\Delta P - \left(b_2\frac{dP_{\textrm{c}}}{d\delta}\Delta \delta \right).
\end{align}
So we can approximate
\begin{align}
a_3 = -\left(a_1 \frac{dP_{\textrm{c}}}{d\delta} + a_2\frac{dT_{\textrm{c}}}{d\delta}\right), \\
b_3 = - b_2 \frac{dP_{\textrm{c}}}{d\delta}.
\end{align}

The critical-line condition \cite{Anisimov_1995b} implies that $(\partial \delta/\partial x)_{T,P} = RT_{\textrm{c}}/x$, therefore
\begin{align}
a_3 = - \frac{x}{\rho_{\textrm{c}}(RT_{\textrm{c}})^2} \left[\frac{dP_{\textrm{c}}}{dx}-\left(\frac{dP}{dT}\right)_\textrm{c,cxc}  \frac{d T_{\textrm{c}}}{dx}\right], \\
b_3 = - \frac{x}{\rho_{\textrm{c}}(RT_{\textrm{c}})^2} \left(\frac{dP_{\textrm{c}}}{dx}\right).
\end{align}
Thus $a_3$ is associated with the Krichevskii parameter in accordance with equation \eqref{eq:Krichevskii}; $b_3$ is associated with the parameter $K_2 = dP_{\textrm{c}}/{dx}$, which plays a secondary role in the behavior of response functions at constant composition.
%We also make the approximation that, for a given $\delta(T,P,x)$, $T_\textrm{c}(x) = T_\textrm{c}(\delta)$ and $P_\textrm{c}(x) = P_\textrm{c}(\delta)$.

We now evaluate the response functions entering equation \eqref{eq:legendre1}. The critical parts of these response functions can be expressed in terms of the scaling susceptibilities, which are defined as follows in the mean-field approximation:
\begin{align}
\chi_1 &= \left(\frac{\partial^2 h_3}{\partial h_1^2}\right)_{h_2}, \\
\chi_2 &= \left(\frac{\partial^2 h_3}{\partial h_2^2}\right)_{h_1} = \phi_1^2 \chi_1, \\
\chi_{12} &= \left(\frac{\partial^2 h_3}{\partial h_1 \partial h_2}\right) = \phi_1 \chi_1. 
\end{align} 
With $b_1 = 0$, the critical parts of the response functions, denoted with a superscript c, read:
\begin{align}
\frac{1}{RT_{\textrm{c}}}\left(\frac{\partial S}{\partial T}\right)_{P,\delta}^{\textrm{c}} &= a_2^2\chi_1, \\
\frac{1}{\rho_{\textrm{c}} RT_{\textrm{c}}}\left(\frac{\partial x}{\partial T}\right)_{P,\delta}^{\textrm{c}} &= a_2a_3\chi_1 + a_2b_3\chi_{12}, \\
\frac{1}{\rho_{\textrm{c}} RT_{\textrm{c}}}\left(\frac{\partial x}{\partial \delta}\right)_{P,T}^{\textrm{c}} &= a_3^2 \chi_1 + 2a_3b_3\chi_{12} + b_3^2 \chi_2.
\end{align}
We approximate the the regular parts of the response functions, denoted by a superscript r, as
\begin{align}
\left(\frac{\partial x}{\partial T}\right)^{\textrm{r}}_{P,\delta} &= 0, \\
\left(\frac{\partial x}{\partial \delta}\right)^{\textrm{r}}_{P,T} &= \frac{x}{RT_{\textrm{c}}}, \\
\left(\frac{\partial S}{\partial T}\right)_{P,\delta}^{\textrm{r}} &= \left(\frac{\partial S}{\partial T}\right)_{P,x}^{\textrm{r}}.
\end{align}
Then, from equation \eqref{eq:legendre1} we have

\begin{equation}
\frac{C_{P,x}}{R}=\frac{C_{P,x}^{\textrm{r}}}{R}+ \hat{a}^{2}\frac{\chi_{1}}{1 + x(\phi_1\hat{\mathcal{L}}+\hat{\mathcal{K}})^2\chi_1}.
\end{equation}

\vspace{3mm}

\subsection{Width of the Two-Phase Region Constant Temperature}

A linear approximation of the coexistence surface,
\begin{equation} \label{eq:cxs}
P-\left(P_{\textrm{c}}^0 + \frac{dP_{\textrm{c}}}{d\delta}\Delta\delta\right) =  \left(\frac{dP}{dT}\right)_{\textrm{c,cxc}}\left( T -\left(T_{\textrm{c}}^0 + \frac{dT_{\textrm{c}}}{d\delta}\Delta\delta\right)\right),
\end{equation}
gives
\begin{align}
\left(\frac{dP}{d\delta}\right)_{T, \textrm{cxc}} &=  \frac{dP_{\textrm{c}}}{d\delta} -  \left(\frac{dP}{dT}\right)_{c,\textrm{cxc}}  \frac{dT_{\textrm{c}}}{d\delta} \\
&\simeq\frac{x}{RT_{\textrm{c}}} \left[\frac{dP_{\textrm{c}}}{dx} -  \left(\frac{dP}{dT}\right)_{\textrm{c,cxc}}  \frac{dT_{\textrm{c}}}{dx} \right] \\
&\simeq \frac{x}{RT_{\textrm{c}}}\mathcal{K},
\end{align}
where here and below $x = x_{\textrm{c}}$, as follows from the critical-line condition (\ref{eq:clc}).
Thus from Eq. \eqref{eq:C-C},
\begin{equation}
\Delta x \simeq -x \mathcal{K}\frac{\Delta V}{RT}.
\end{equation}
To a first approximation, the width of the two-phase region can be estimated as
\begin{equation}
\Delta T \simeq  \Delta x \left(\frac{dT}{dx}\right)_{P,\overline{x}},
\end{equation}
where 
$(dT/dx)_{P,\overline{x}}$ is the slope of the line of symmetry (along the average concentration $\overline{x}$) of the two-phase region in a $(T,x)$ plane.  The approximate slope of this line is
\begin{equation}
\left(\frac{dT}{dx}\right)_{P,\overline{x}} = \frac{x}{RT_{\textrm{c}}}\left(\frac{dT}{d\delta}\right)_{P,\textrm{cxc}}.
\end{equation}
From Eq. \eqref{eq:cxs},
\begin{equation}
\left(\frac{dT}{dx}\right)_{P,\overline{x}} \simeq \mathcal{K} \left(\frac{dT}{dP}\right)_{\textrm{c,cxc}}.
\end{equation}
Therefore,
\begin{equation}
\Delta T \simeq x\mathcal{K}^{2}\frac{\Delta V}{RT}\left( \frac{dT}{dP}\right)_{\mathrm{c,cxc}}. 
\end{equation}

\bibliography{../../PhysRefsv2}

\end{document}